\documentclass[sigconf,urlbreakonhyphens=false]{acmart}

\AtBeginDocument{%
  }

\copyrightyear{2025}
\acmYear{2025}
\setcopyright{rightsretained}
\acmConference[SUI '25]{ACM Symposium on Spatial User Interaction}{November 10--11, 2025}{Montreal, QC, Canada}
\acmBooktitle{ACM Symposium on Spatial User Interaction (SUI '25), November 10--11, 2025, Montreal, QC, Canada}
\acmDOI{10.1145/3694907.3765954}
\acmISBN{979-8-4007-1259-3/2025/11}

\usepackage{etoolbox}
\makeatletter

\author@bx@wd=\dimexpr(\textwidth-2\author@bx@sep)/3\relax

\patchcmd{\@mkauthors@iii}{%
  \or 
  \divide\author@bx@wd by \num@authorgroups\relax
}{%
  \or 
  \author@bx@wd=\dimexpr(\textwidth-2\author@bx@sep)/3\relax
}{}{
  \typeout{Failed to patch 3-author width calculation}
}

\patchcmd{\@mkauthors@iii}{%
  \advance\author@bx@wd by -\author@bx@sep\relax
}{%
}{}{
  \typeout{Failed to skip final width adjustment}
}

\patchcmd{\@mkauthors@iii}{%
  \lineskip=1pc\relax\centering\hspace*{-1em}%
}{%
  \lineskip=1pc\relax\centering
}{}{
  \typeout{Failed to adjust centering}
}

\makeatother

\begin{document}

\title[Instant Skinned Gaussian Avatars]{Instant Skinned Gaussian Avatars for Web, Mobile and VR Applications}

\author{Naruya Kondo}
\orcid{0000-0002-9694-4676}

\affiliation{
  \institution{University of Tsukuba}
  \city{Tsukuba}
  \state{Ibaraki}
  \country{Japan}
}
  \email{n-kondo@digitalnature.slis.tsukuba.ac.jp}

\author{Yuto Asano}
\affiliation{
  \institution{University of Tsukuba}
  \city{Tsukuba}
  \state{Ibaraki}
  \country{Japan}
}
  \email{yuto.asano@digitalnature.slis.tsukuba.ac.jp}

\author{Yoichi Ochiai}
\affiliation{
  \institution{University of Tsukuba}
  \city{Tsukuba}
  \state{Ibaraki}
  \country{Japan}
}
  \email{wizard@slis.tsukuba.ac.jp}

\renewcommand{\shortauthors}{Kondo et al.}

\begin{abstract}
We present Instant Skinned Gaussian Avatars, a real-time and cross-platform 3D avatar system. Many approaches have been proposed to animate Gaussian Splatting, but they often require camera arrays, long preprocessing times, or high-end GPUs. Some methods attempt to convert Gaussian Splatting into mesh-based representations, achieving lightweight performance but sacrificing visual fidelity. In contrast, our system efficiently animates Gaussian Splatting by leveraging parallel splat-wise processing to dynamically follow the underlying skinned mesh in real time while preserving high visual fidelity. From smartphone-based 3D scanning to on-device preprocessing, the entire process takes just around five minutes, with the avatar generation step itself completed in only about 30 seconds. Our system enables users to instantly transform their real-world appearance into a 3D avatar, making it ideal for seamless integration with social media and metaverse applications.

\end{abstract}

\ccsdesc[500]{Computing methodologies~Shape modeling}
\ccsdesc[500]{Computing methodologies~Animation}
\ccsdesc[300]{Computing methodologies~Virtual reality}
\ccsdesc[300]{Computing methodologies~Mixed / augmented reality}

\keywords{Gaussian Splatting, Avatar}


\begin{teaserfigure}
  \centering
  \includegraphics[width=0.9\textwidth]{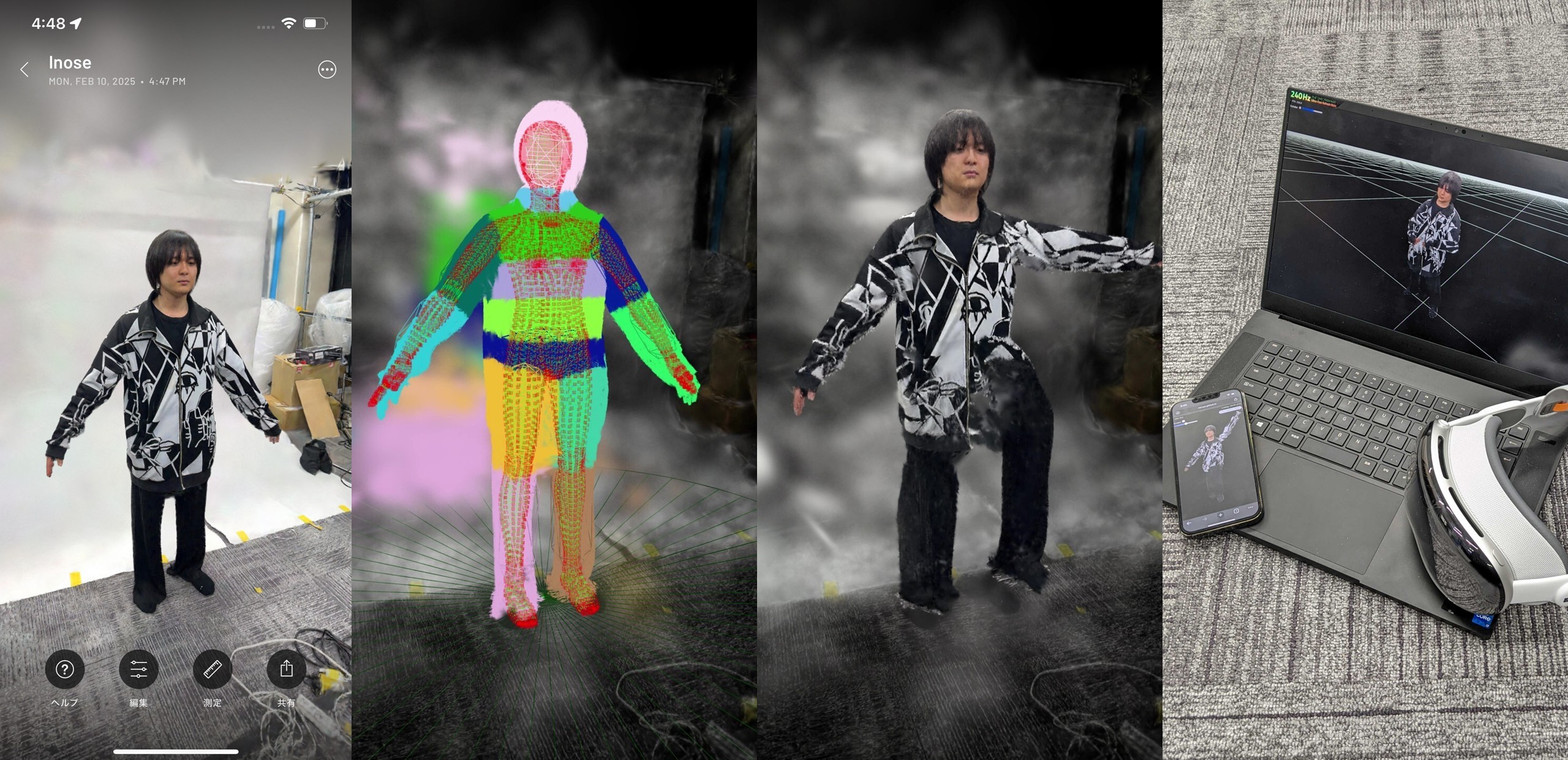}
  \caption{Instant Skinned Gaussian Avatar generates Gaussian Splatting-based photo-realistic 3D avatars in just 5 minutes. From scanning to avatar creation, the process is completed on a smartphone, and thanks to our WebXR-based approach, these avatars can be seamlessly deployed across VR, AR, and web platforms. Figure Caption: (a) A static Gaussian Splatting model of the scene reconstructed using the off-the-shelf Scaniverse application. (b) Our system moves the Gaussian splats by binding them to the vertices of a background mesh. Once processed, the human avatar starts animating within the original reconstruction space. (d) The resulting avatar can be easily integrated into metaverse applications.}
  \label{fig:teaser}
\end{teaserfigure}

\maketitle

\section{Introduction}

Creating 3D avatars of human subjects has traditionally relied on either manual modeling or photogrammetry \cite{schoenberger2016mvs} pipelines that automatically reconstruct 3D geometry from 2D data. Recently, Gaussian splatting \cite{kerbl3Dgaussians} has emerged as a powerful technique that enables high-fidelity scene reconstruction and real-time rendering, prompting various studies on its application to 3D avatar creation \cite{li2024animatablegaussians, qian20233dgsavatar, GauHuman, hugs, zielonka25dega, Pang_2024_CVPR, shao2025degas, hu2024gaussianavatar}. However, many existing methods still require camera arrays, lengthy preprocessing, or high-end GPUs, and attempts to convert splats into mesh-based representations often compromise visual fidelity. Consequently, 3D avatar creation remains inaccessible for everyday users. ExAvatar \cite{moon2024exavatar}, can generate 3D avatars from a single 10-second video but requires two to three hours of preprocessing and relies on SMPL meshes that reduce expressive 3D detail. In this study, we propose Instant Skinned Gaussian Avatars. Using only a single camera video as the data source, it can be incorporated into applications with only about 5 minutes total processing, of which avatar generation itself takes just around 30 seconds. This system simply makes gaussian splats follow a mesh that moves in the background while maintaining relative transformations. This process is normally very computationally intensive, but by using per-splat parallel processing, we have achieved real-time processing without losing expressive power.

Owing to its immediacy and mobile compatibility, our approach has the potential to make realistic avatars more accessible to a wider public. In particular, it is expected to be effective in virtual environments where formality is required, as well as in digital twin settings that demand photorealism.

\section{Instant Skinned Gaussian Avatars}

\subsection{Overview}
In our system, we simply animate a background 3D mesh and have the gaussian splats follow the mesh's vertices. During preprocessing, splats are assigned to mesh vertices, and their relative transformation are stored. Once this data is saved, you can instantly use it in your applications without further preprocessing. At runtime, we animate the background 3D mesh, update the gaussian splats in parallel, and resort all gaussian splats every frame based on the viewer’s perspective.

\subsection{Preprocessing}
The preprocessing workflow consists of the following five steps: (1) we first use a pre-existing scanning application called ``Scaniverse'' \cite{scaniverse} to capture the subject in Gaussian Splatting format; to streamline subsequent preprocessing, we ask the subject to assume an A-pose. (2) We then remove any points that do not belong to the subject; since the subject is always near the center of the Gaussian Splatting scene, we eliminate unwanted splats both horizontally and vertically through simple rule-based filtering, and then we normalize the splat's position and scale. (3) Using pose estimation, we infer the subject’s front-facing direction and limb angles, and position a background 3D mesh at the same place, pose, and scale. (4) For each splat, we select the nearest mesh vertex through a nearest-neighbor search, calculate its relative transformation. (5) Finally, our system output the subject’s pose, scale, the nearest-vertex indices, and the corresponding relative transformations.

\subsection{Animation}
 We perform three steps each frame: (1) animate the background mesh, (2) update the positions and orientations of the Gaussian splats, and (3) sort the splats according to the viewer’s perspective. Because our system uses a standard skinned mesh in the background, any existing animations can be used as-is. Moreover, by having the splats follow the mesh vertices, the original skinning deformations around joints—such as stretching and compressing—are effectively preserved on the Gaussian splats. One notable issue in animating Gaussian splats is the high computational cost of sorting them from the viewer’s perspective. In static Gaussian splatting, position data is cached on the GPU, allowing large splat counts to be sorted with high efficiency. However, in dynamic Gaussian splatting, the positions must be updated each frame, causing a severe drop in performance if handled naively. To mitigate this, we group splats and sort them at the group level instead. There is a trade-off between the number of groups and the required hardware capabilities; to ensure mobile compatibility, we opted for bone-level grouping based on the background mesh.

\subsection{Implementation}

Our system, excluding the scanning process using Scaniverse, is entirely built as a browser application using JavaScript and Three.js. As a result, it is readily accessible on mobile devices, VR headsets, and other platforms. For the background mesh, we use a single VRM-format 3D avatar mesh with a neutral body shape and 32k polygons. This should enable seamless integration with VRM-based applications, such as motion capture-driven avatar experiences and VRM-compatible games. The preprocessing steps are fully automated. The 3D reconstruction in Scaniverse takes approximately one minute on an iPhone 13 Pro, while the remaining preprocessing (avatar generation) steps are now completed in around 30 seconds on the same device. Including the scanning process, an avatar can be fully created in about five minutes. During animation, our system runs at 40–50 fps on an iPhone 13 Pro and reaches 240 fps (capped by the display’s refresh rate) on a laptop equipped with an NVIDIA GeForce RTX 3060 Laptop GPU.



\bibliographystyle{ACM-Reference-Format}
\bibliography{base}


\end{document}